# Importance of Biologically Active Aurora-like Ultraviolet Emission: Stochastic Irradiation of Earth and Mars by Flares and Explosions

David S. Smith[1], John Scalo[1] and J. Craig Wheeler[1]

[1]University of Texas at Austin

(*Author for correspondence:  John Scalo; email: parrot@astro.as.utexas.edu; telephone: 512-478-2748; fax: 512-471-6016)

## Abstract

Habitable planets will be subject to intense sources of ionizing radiation and fast particles from a variety of sources—from the host star to distant explosions—on a variety of timescales.  Monte Carlo calculations of high-energy irradiation suggest that the surfaces of terrestrial-like planets with thick atmospheres (column densities greater than about 100 g cm$^{-2}$) are well protected from directly incident X-rays and $\gamma$-rays, but we find that sizeable fractions of incident ionizing radiation from astrophysical sources can be redistributed to biologically and chemically important ultraviolet wavelengths, a significant fraction of which can reach the surface. This redistribution is mediated by secondary electrons, resulting from Compton scattering and X-ray photoabsorption, the energies of which are low enough to excite and ionize atmospheric molecules and atoms, resulting in a rich aurora-like spectrum.  We calculate the fraction of energy redistributed into biologically and chemically important wavelength regions for spectra characteristic of stellar flares and supernovae using a Monte-Carlo transport code and then estimate the fraction of this energy that is transmitted from the atmospheric altitudes of redistribution to the surface for a few illustrative cases.  For atmospheric models corresponding to the Archean Earth, we assume no significant ultraviolet absorbers, only Rayleigh scattering, and find that the fraction of incident ionizing radiation that is received at the surface in the form of redistributed ultraviolet in the biologically relevant 200-320 nm region (UV-C and UV-B bands) can be up to 4%.  On the present-day Earth with its ultraviolet ozone shield, this fraction is found to be 0.2%.  Both values are many orders of magnitude higher than the fraction of direct ionizing radiation reaching the surface. This result implies that planetary organisms will be subject to mutationally significant, if intermittent, fluences of UV-B and harder radiation even in the presence of a narrow-band ultraviolet shield like ozone. We also calculate the surficial transmitted fraction of ionizing radiation and redistributed ultraviolet radiation for two illustrative evolving Mars atmospheres whose initial surface pressures were 1 bar.  We discuss the frequency with which redistributed ultraviolet flux from parent star flares exceeds the parent star ultraviolet flux at the planetary surface.  We find that the redistributed ultraviolet from parent star flares is probably a fairly rare intermittent event for habitable zone planets orbiting solar-type stars except when they are young, but should completely dominate the direct steady ultraviolet radiation from the parent star for planets orbiting all stars less massive than about 0.5 solar masses.  Our results suggest that coding organisms on such planets (and on the early Earth) may evolve very differently than on contemporary Earth, with diversity and evolutionary rate controlled by a stochastically varying mutation rate and frequent hypermutation episodes.

## 1.  Introduction

The origin and evolution of life on Earth and presumably other planets is constrained and directed in part by the surface ultraviolet (UV) radiation environment. UV effects include DNA damage and associated repair pathways, the development of cell membranes, and many other cellular phenomena (see Rothschild 1999 for a thorough review). Astronomical ionizing radiation (X-rays and $\gamma$-rays), for example from solar flares, has been thought to be important only insofar as it affects phenomena associated with the ionosphere.  Previous models and observations for estimating the ionization profiles (Sharma et al. 1972, Brown 1973, Kasturirangan et al. 1976, O'Mongain and Baird 1976, Fishman and Inan 1988, Inan et al. 1999) or ozone chemistry (Gehrels et al. 2003) in the Earth's atmosphere due to X-ray and $\gamma$-ray events have not been concerned with the possibility of direct biological effects at planetary surfaces.  Other planets and satellites in the solar system and exoplanetary systems with thin or no atmospheres that directly transmit substantial fractions of incident ionizing radiation are only marginally relevant for the question of extraterrestrial life, and discussions of biological effects have been restricted to bodies like present-day Mars and Europa, or the space environment (e.g., Baumstark-Khan and Facius 2000).  Such bodies, while interesting as survival tests, have, at the present, too small an atmospheric pressure to support a liquid ocean unless heated by tidal effects, as may be the case for Europa.  Organisms inhabiting exoplanets with thick atmospheres like the Earth's have traditionally been assumed to be extremely well protected from ionizing radiation, since the optical depth at, for example, 1 MeV is about 65 for the Earth.  Thus the only biologically relevant stellar radiation has been assumed to be the ultraviolet photons that make their way past whatever UV shield (e.g., ozone) exists (see, e.g., Cockell 2000a, b, 2002 and references therein).  The purpose of the present paper is to show that a substantial fraction of incident *ionizing* radiation can be transformed to ultraviolet radiation by a process analogous to aurorae involving secondary electron impact excitation of atmospheric molecules, and to estimate the fraction of this "redistributed" UV energy that reaches the surface.

Planets orbiting the Sun and other stars are occasionally subjected to large ionizing fluxes from astronomical sources.  The disturbance of the Earth's upper atmosphere due to solar activity is a frequent occurrence, and on longer timescales, more intense flares and associated phenomena must surely occur.  The surfaces and atmospheres of satellites of Jupiter and Saturn are strongly affected by particles and photons from those planets.  Habitable planets can be subjected to far more intense sources of irradiation by ionizing photons and particles on a variety of timescales. All conventional habitable zone (HZ, continuous surface liquid water—see Brack 2000 for a discussion of the rationale behind this requirement, Kasting et al. 1993 and Franck et al. 2000 for calculations of the habitable zone width) planets orbiting stars with masses like the Sun's or less will be subject to stellar flares with fluxes that depend on time scale and stellar spectral type. For example, HZ planets orbiting low-mass, main sequence stars (spectral type dMe) will experience a fluence of $10^7$-$10^9$ erg cm$^{-2}$ roughly once per 100 hours above the atmosphere for several billion years, with larger fluences at longer intervals (see, for example, the flare energy distribution found by Güdel et al. 2003). Even the steady coronal X-ray fluxes of these stars will be on average $10^2$-$10^4$ times larger than the Earth's exposure to the sun's coronal X-rays (see A. Andreeshchev, J. Scalo, and D. S. Smith 2003, in preparation). That the most numerous stars in the Galaxy are very low-mass stars (see Chabrier 2001 and references therein) underscores the importance of ionizing radiation for the evolution of life on exoplanetary systems.

The frequency with which stochastic irradiation affects planetary atmospheric chemistry and biological activity through direct mutational enhancement or sterilization has been estimated by Scalo and Wheeler (2002) and by Scalo, Wheeler, and Williams (2003). $\gamma$-ray bursts subject all planets to short-duration (~10 sec) hard photons, resulting in a biologically significant (hypermutational or lethal) fluence for eukaryotic-like organisms every 100–500 Gyr (Scalo and Wheeler 2002), with smaller frequencies for prokaryotes because of their generally larger mutation doubling or lethal doses. Being brief, $\gamma$-ray bursts should not affect evolution directly by mutation, but may partially sterilize surface and upper oceanic organisms and indirectly affect evolution through changes in atmospheric chemistry and by niche reorganization. Type II and Ibc supernova $\gamma$-ray lines associated with decay of $^{56}$Co and other species should irradiate any planet in the Galaxy with chemically and biologically significant fluences for a duration of weeks at a rate of $10^3$-$10^4$ Gyr$^{-1}$, depending on Galactic location. Cosmic-rays arriving later may also be potent sources of shower $\gamma$-rays as well as fast particles. The supernova events may persist long enough for fixation of mutations, neutral or adaptive, in small populations of organisms with relatively short generation times in the case of decay $\gamma$-rays, or in any terrestrial organisms in the case of cosmic-rays, whose irradiation duration may be $10^2$-$10^5$ yr. The potential evolutionary importance of the long duration of the follow-up cosmic-ray particles in supernova explosions was pointed out by Sagan and Shklovskii (1966).

The radiation environment of most planets, in the solar system and elsewhere, will thus be highly stochastic on a variety of time scales. The biological effects depend strongly on the fraction of the incident radiation that can reach a given level of the atmosphere or reach the surface. For this reason, we have studied the transfer of radiation and energy deposition of $\gamma$- and X-ray radiation through planetary atmospheres with a variety of total column densities, from atmospheres thinner than that of Mars to those thicker than that of Earth. We have employed Monte Carlo transport calculations to study a broad range of atmospheric column densities and the dependence on parameters such as incident energy and spectra, angle of incidence, and planetary surface gravity. We wished to substantially improve on estimates of planetary surficial radiation fluxes based on exponential attenuation or single scattering approximations. We have in the process uncovered a new aspect of the physics that results in transmitted fractions of biologically important radiation that is many orders of magnitude larger than what would follow from an attenuation model. We follow the Compton scattering and X-ray photoabsorption in full detail and offer an approximate estimate of the redistribution of X-rays to surface UV radiation by excitation of molecular energy levels by secondary electrons emitted along the path of primary electrons produced by photoabsorption and Compton scattering. We outline our radiative transfer calculations in Section 2, and present our results applied to the Archean Earth, present-day Earth, and the atmospheric evolution of Mars as proxies for generic habitable exoplanets in Sections 3.A, 3.B, and 3.C, respectively. The environments in which the redistributed UV radiation will be important in comparison with the parent star's UV radiation are discussed in Section 4. Our conclusions are summarized in Section 5.

## 2. Radiative Transfer and UV Redistribution

To evaluate the effects of bursts of ionizing radiation, we have modeled the propagation of incident radiation through atmospheres characteristic of an evolving Mars, the present-day Earth, and Archean Earth. We assume for simplicity that the Archean Earth had 1 bar of pure nitrogen only and that the present-day Earth has the contemporary concentrations of $N_2$, $O_2$ and $O_3$. The unique case of Mars is discussed below in Section 3.C. The calculations were performed using a weighted Monte Carlo approach that accurately handles the complicated angular and energy dependence of the Compton cross section (Hammersley and Handscomb 1979, Kalos and Whitlock 1986, Watson and Henney 2001). Photons are initialized at the top of the atmosphere heading downward, with energies distributed according to a specified incident spectrum, and then propagated by choosing a random optical depth, calculating the distance corresponding to that optical depth given the cross section at the photon's energy, and then scattered. Because the optical depths of the exoplanet atmospheres of interest range from ~0.1 to 100, a weighting procedure was used to minimize fluctuations due to small number statistics (Watson and Henney 2001). Details of the radiative transfer calculations, their testing, and results as a function of column density and other parameters are presented in Smith, Scalo, and Wheeler (2003). For atmospheres with column densities as thick as the Earth's, it was necessary to initialize the calculation with about a million photons. Because of the weighting procedure used (which assigns partial outcomes for various interactions) it is not meaningful to state how many photons were still propagating at the surface. However we made sure that the fluctuations in the resulting spectra were insignificant, tested the code on several test problems (e.g. Comptonization of photons off a population of cold electrons, which should give a Gaussian distribution, as shown by Xu, Ross, and McCray 1991), and verified that we obtain identical results when running models that differ only in the sampling of initial seed random numbers.

Compton scattering was treated in three dimensions using the Klein-Nishina cross section and angular distribution of scattered photons given by Lingenfelter and Rothschild (2000), and the effects of photoelectric absorption were included using the cross section formula given by Setlow and Pollard (1962, p. 316). We find that the latter is an excellent fit, for energies larger than the K edge, to the element-by-element cross sections measured and tabulated by Henke, Gullikson, and Davis (1993). More complex effects at lower energies (e.g., collective effects, in which the electrons cannot be treated as independent particles) are unimportant for the energies encountered here. Pair production is unimportant for the energies studied here and so was neglected; it becomes important above energies of around 5 MeV for the atmospheric compositions of interest. The incident photon energy spectrum, the angle of incidence, the atmospheric column density and the atmospheric scale height were specified as model input parameters. Because of the large photon energies, the transfer of ionizing radiation is independent of the adopted chemical composition of the atmosphere, except for the number of electrons per molecule; the photons essentially "see" free electrons.

For the results presented here, we adopt an atmospheric scale height of 8 km, similar to that of the Earth, and normal incidence of incoming photons. The results are not very sensitive to these parameters (Smith et al. 2003). We use two illustrative types of incident spectra, monochromatic photons of 0.25, 0.5, 1, and 2 MeV as representative of supernova radioactive decay photons, and an exponential spectrum for stellar flares, assuming a flare spectrum will be similar to the emission from a hot plasma (Tucker and Koren 1971). We show elsewhere that the results for a continuous spectrum of $\gamma$-ray or hard X-ray photons are very similar to the results for monochromatic incident spectra, so the 0.25 and 0.5 MeV cases can be taken as representing $\gamma$-ray bursts (see Band et al. 1993, Band 2001 for spectra) as well as supernova decay photons. The adopted stellar flare spectra have the form $F(E) \propto \exp(-E/E_p)$, where $F(E)$ is the flux per unit energy, $E$ is the energy, and $E_p$ is parameter which adjusts the average incident energy; we chose average incident energies of 1 and 10 keV. Interestingly, it will be seen that the amount of redistributed UV that reaches the surface is relatively insensitive to the form of the incident spectrum.

Each photoabsorption of an X-ray or Compton scattering of a $\gamma$-ray results in a free electron of energy much larger than the ionization potential of the target molecules. These primary electrons lose energy by collisional ionizations, each of which results in a secondary electron of average energy 35 eV for $N_2$ (and not much different for a $CO_2$ atmosphere). We treat the secondary electron spectrum as a $\delta$-function at 35 eV. This determines the number of ionizations each primary electron generates. Since the atmosphere is essentially neutral, very little of the secondary electron energy will be thermalized by Coulomb collisions, so we ignore this mode of energy dissipation (see Smith et al. 2003 for a detailed justification). Instead, the secondary electrons ionize and excite molecular levels. Subsequent radiative de-excitation yields photons with a broad range of energies. This process is similar to that which produces aurorae (e.g., Chamberlain 1961). Radiation from secondary electron impact excitation has been studied in the context of cosmic-ray-generated UV radiation in interstellar clouds (Prasad and Tarafdar 1983, Gredel et al. 1989) and the UV and blue emission from accretion disks around compact objects, especially active galactic nuclei (see Ross 1979, Kallman and McCray 1982, and Ross and Fabian 1993 for detailed calculations). Although these previous calculations were dominated by atomic or molecular hydrogen instead of the molecules expected in an exoplanetary atmosphere, it is significant that the fraction of energy redistributed to the UV was found to be large in these papers. A diagram illustrating this electron-mediated redistribution process is shown in Figure 1, where the thick lines indicate the connections we found to be most important and include in the present calculations.

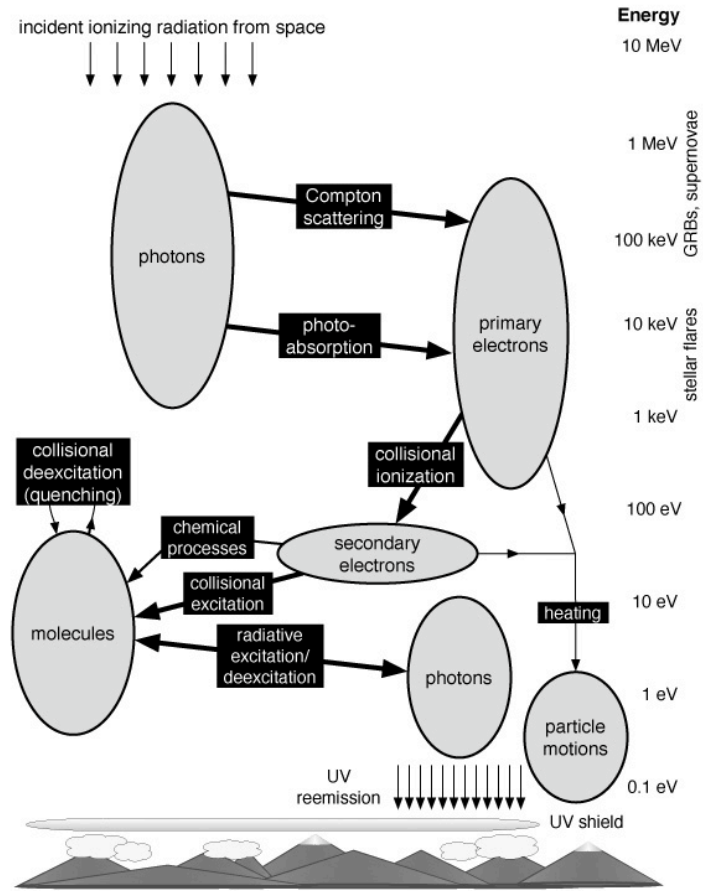

**Figure 1**: Schematic illustration of the physical processes that mediate the redistribution of the incident energy from ionizing gamma- and X-ray photons (arrows at top left) to UV aurora-like reemission (arrows at bottom right). Thick lines indicate processes found to be important and/or included in the present study. Chemical processes are not included, not because they will not be affected by the irradiation events, but because the redistribution process is relatively insensitive to the detailed atmospheric abundances. "Heating" refers to sharing of primary and secondary electron kinetic energy with thermal energies of other particles through Coulomb or neutral-neutral collisions. "Quenching," or collisional deexcitation of the levels excited by secondary electrons, was found to be unimportant for lifetimes and deexcitation cross sections representative of various $N_2$ electronic levels associated with transitions leading to ultraviolet emission. A detailed discussion of heating and quenching is presented in Smith et al. (2003).

Our adopted redistribution procedure contains the essential physics in parameterized form, guided by observed auroral spectra. For simplicity, we assume the secondary electron energy is completely and locally redistributed as a spectrum $F(\lambda)$ at each layer in a pure $N_2$ atmosphere. The electron excitation cross sections as a function of energy for $N_2$ have broad maxima around 10 to 50 eV, so we expect these levels to be well populated, since the average secondary electron energy is 35 eV. This suggests that the Vegard-Kaplan and 2nd positive electronic/vibrational transitions in $N_2$ and the Lyman-Birge-Hopfield and 1st negative bands of $N_2^+$ will be sources of strong UV reradiation, as they are in auroral spectra (Jones 1974). The emitted energy is very roughly distributed equally among the many lines as in aurorae, so we assume the deposited X-ray energy is uniformly reradiated between the wavelengths corresponding to $N_2$ and $O_2$ photoionization continua (~ 100 nm) and the extent of significant lines in the $N_2^+$ 1st negative transition (~ 600 nm).

The Monte Carlo calculations give the energy deposited by X-ray photoabsorption and Compton recoil at each layer. This energy, spread as a $1/\lambda$ spectrum (to mimic the form of auroral spectra) from 100-600 nm, is emitted isotropically and propagated to the ground using either exponential attenuation (Beer's Law) for $O_2/O_3$ absorption (similar to present-day Earth), or the two-stream scattering approximation (see Thomas and Stamnes 1999, Ch. 7) in the case that Rayleigh scattering provides the only UV opacity (assumed similar to Archean Earth). Finally, we sum the contributions from each layer and integrate the spectrum at the ground from 200 to 320 nm, since this is the region of greatest biological activity for terrestrial organisms. This procedure could easily be generalized to include $CO_2$ in the atmospheric composition, but since $CO_2$ absorbs photons effectively below only about 190 nm, it would not affect our estimate of the biologically effective surficial radiation. The surface transmitted energy fractions we calculate in this region are upper limits, since we neglect aerosols and molecular absorbers other than $N_2$, and $O_2/O_3$ in the contemporary Earth model. However the actual fractions for the Archean case may not be much smaller, given recent evidence for a lack of an Archean UV shield based on sulfur isotope fractionation (Farquhar et al. 2002, Wiechert 2002).

The ratio of the energy that reaches the ground in the 200-320 nm region to the total incident ionizing energy will be referred to as the "UV transmittance," or "redistributed UV fraction," with the understanding that this fraction refers only to the 200-320 nm region. The details of the sensitivity to the choice of the upper wavelength cutoff are discussed elsewhere (Smith et al. 2003).

# 3. Results

## A. Archean Earth

The Archean Earth presents perhaps the simplest case of a habitable planet, with respect to the atmospheric radiative transfer. In particular, before the rise of oxygen, the atmosphere lacked the current ozone shield, and so (neglecting aerosols or other potential UV shields; see Cockell 2002 and references therein) the atmosphere may have been relatively transparent to UV in the biologically effective 200-320 nm range. With no significant molecular absorption, the re-emission due to UV redistribution is subject only to Rayleigh scattering. Using a modified two-stream approximation for a pure scattering atmosphere, we propagated the reemitted UV through an Earth-like atmosphere of pure $N_2$ and a column density of 1024 g cm$^{-2}$. Although we recognize that there is considerable uncertainty concerning UV screening in the Archean atmosphere (e.g., Levy and Miller 1998, Cockell 2002), recent evidence concerning mass-independent fractionation effects in Archean sulfides (Farquhar et al. 2002) suggest the absence of a significant UV shield during this period (Wiechert 2002), so our assumed Archean atmosphere may not be so extreme. Table 1 shows the resulting fractions of the incident energy in both ionizing radiation and biologically effective UV received at the surface for three different incident spectra. As the incident ionizing radiation is efficiently blocked by atmospheres as thick as the Earth's, the surface fluences are completely dominated by the redistributed UV, showing the potential importance of this physical process in determining the radiation environment on the early Earth or terrestrial-like exoplanets without UV shields.

The transmittance of the direct ionizing radiation rapidly decreases by orders of magnitude with decreasing energy because the photoabsorption cross section, which scales as the inverse cube of the energy, is rapidly increasing. Thus the fraction of the incident ionizing radiation received at the ground is ridiculously small for the 1 and 10 keV average energy exponential spectra, which are characteristic of stellar flares. Including the secondary electron-mediated redistribution of the absorbed ionizing radiation to biologically effective UV raises the surficial fluxes by *many* orders of magnitude. As a concrete example, an estimate of the surface environment of the Archean Earth under irradiation by a nearby supernova's radioactive decay $\gamma$-ray lines would be too low by almost 30 orders of magnitude without including the UV redistribution. A similar estimate of the Archean Earth under irradiation by a solar superflare would be wrong by 100 orders of magnitude.

| Spectrum Type | Ionizing Transmittance | UV Transmittance (200-320 nm) |
|---|---|---|
| 1 keV $\langle E \rangle$ exponential | ~0 | 3.9 x 10$^{-2}$ |
| 10 keV $\langle E \rangle$ exponential | 1.9 x 10$^{-110}$ | 4.0 x 10$^{-2}$ |
| 1 MeV monoenergetic | 6.4 x 10$^{-29}$ | 4.4 x 10$^{-2}$ |

**Table 1**—Fractional surficial fluences on the Archean Earth for three representative types of incident X-ray and γ-ray spectra. In each case, the fraction of the incident radiation reaching the ground is much larger when redistribution of the ionizing radiation to the biologically effective 200-320 nm region is included.

Notice that the fraction of the incident fluence redistributed to UV and reaching the surface is almost independent of the value of the mean energy of the incident photons. This occurs because the number of secondary electrons available for molecule excitation only depends on the *total* energy of primary electrons, which in turn depends on the *total* fluence of ionizing radiation.

## B. Present-day Earth

The present-day Earth is an interesting prototype of a simple atmosphere with a single important molecular UV absorber that peaks in the 200-320 nm region. Although any planet with a thick atmosphere like Earth is well protected from $\gamma$-rays and X-rays, the planet may or may not be protected from UV irradiation. Currently, the most important UV absorber on Earth in the biologically effective wavelength region is ozone, a byproduct of solar UV photolysis of oxygen in the lower atmosphere. The wavelength-dependent cross section of ozone to UV absorption peaks at around 260 nm, rapidly falling to higher and lower wavelengths and becomes small outside the range 210-290 nm. Not all energy in the reemitted aurora-like spectrum lies in the range of significant ozone absorption, however, and a significant amount does "leak" through the sides of the ozone absorption peak in regions where biological effects are known to be significant but ozone absorption is minimal. The choice of the upper wavelength for biological effectiveness sensitively affects the results because of the rapidly declining ozone cross section in the wavelength range 300-350 nm. While a variety of biological effects in certain organisms have been observed up to wavelengths of 350 nm and larger (e.g., Jagger 1985, Nilsson 1996), we conservatively placed the upper limit in our calculations at 320 nm, given that we are interested only in gauging the significance of the redistributed UV on a generic organism.

Because the altitudes of energy deposition in the case of incident $\gamma$-radiation are relatively small and overlap the Earth's ozone layer, we assumed the average vertical dependence of the ozone mixing ratio from Brasseur, Orlando, and Tyndall (1999, Appendix C). We assumed that the $O_2$ mixing fraction is constant. The wavelength dependence of the $O_2$ and $O_3$ absorption cross sections were digitized versions of the graphical results presented by Yung and DeMore (1999).

According to our calculations, the UV transmittance in the presence of the modern ozone shield is nearly identical for any incident spectrum within the constraints of the sources addressed in this work, for the same reason as given above in the case of no UV shield. For the three spectra listed in Table 2, the ionizing transmittance is of course identical to that of the Archean Earth, but the UV transmittance in each case on the present-day Earth drops to $2.1 \times 10^{-3}$. This is somewhat below the value for the early Earth, but it is still orders of magnitude above the fraction of direct ionizing radiation transmitted. The surprisingly large UV transmittance leads us to conclude that the UV redistribution mechanism is extremely important even when the reemission is attenuated through molecular absorption by ozone, and that the Earth is presently unprotected from the types of astrophysical events discussed in Sec. 1.

| Spectrum Type | Ionizing Transmittance | UV Transmittance (200-320 nm) |
|---|---|---|
| 1 keV $\langle E \rangle$ exponential | ~0 | $2.1 \times 10^{-3}$ |
| 10 keV $\langle E \rangle$ exponential | $1.9 \times 10^{-110}$ | $2.1 \times 10^{-3}$ |
| 1 MeV monoenergetic | $6.4 \times 10^{-29}$ | $2.1 \times 10^{-3}$ |

**Table 2**—Fractional surficial fluences on the present-day Earth for three representative types of incident X-ray and $\gamma$-ray spectra. In each case, the fraction of the incident radiation reaching the ground is much larger when redistribution of the ionizing radiation to the biologically effective 200-320 nm region is included, even in the presence of an efficient molecular absorber such as ozone.

## C. Mars as an Evolving Example

Mars presents an interesting application of the present calculations because its atmospheric column density has decreased with time and is now thin enough for a significant fraction of the incident ionizing radiation to directly reach the surface. Although the present mean surface pressure is about 6.4 mbar (Lodders and Fegley 1998), there are several lines of evidence, primarily from outgassing model constraints and geomorphic evidence for early liquid water, that Mars had a thick (~ 1 bar) $CO_2$ atmosphere about 4 Gyr ago (e.g., Carr 1999, Jakosky and Phillips 2001). Depletion of $CO_2$ by incorporation into the regolith, ice caps, and carbonates, as well as atmospheric escape by sputtering and other non-thermal and thermal processes, have been examined in a large number of studies. There has been little consensus about the evolution of the total pressure (cf. Haberle et al.1994 and Leblanc and Johnson 2001). Rather than a continuous pressure decrease, some models even yield a "climate collapse," involving an almost discontinuous reduction in $CO_2$ due to ice formation at age 2.5-3.5 Gyr, indicating that Mars has had a thin atmosphere for ~ 1 Gyr (Haberle et al. 1994). Cockell (2000a) has pointed out the possibility of an "ultraviolet crisis" resulting from climate collapse, considering only the steady UV from the Sun. In any case, Mars should have been subjected to brief optically thin exposures to sterilizing γ-rays and hard X-rays from solar flares, supernovae, and γ-ray bursts many times during the past Gyr. Similarly, if Mars began with a thick atmosphere, then its early evolution would have been punctuated with bursts of UV representing redistributed X-rays from the same astronomical sources. A study of the attenuation of *steady* galactic cosmic-ray and solar UV radiation at the Martian surface as a function of time has been given by Molina-Cuberos et al. (2001). Cordoba-Jabonero et al. (2003) present an accurate calculation of the radiative transfer of steady solar UV in the presence different column densities of volcanic ashes and $SO_2$ UV shields on present-day Mars, based on a detailed photochemical model for Mars' atmosphere.

Given the uncertainties involved in the Martian atmospheric history, we chose to calculate the fractional transmission of ionizing radiation as a function of time for simple models with a pure $CO_2$ atmospheric composition. We examined both power law and exponential models in which the $CO_2$ pressure decreased from 1 bar at 4 Gyr ago (after intense planetesimal bombardment subsided) to 6.4 mbar at present. The exponential version of this decreasing pressure is very similar to the moderate sputtering loss history favored by Luhmann et al. (1992) and the more detailed three-dimensional Monte Carlo sputtering calculations by Leblanc and Johnson (2001). We neglect any UV shield other than Rayleigh scattering, since the composition and altitude dependence of such shields depends on the treatment of a complex photochemical model and assumptions about geological activity (see Cordoba-Jabonero et al. 2003) which would have to be extended to earlier epochs. This uncertainty means that, as in the case of Archean Earth calculations, the fraction of redistributed energy that is transmitted to the surface calculated here is an upper limit.

Figure 2 shows the fractional fluence arriving at the Martian surface as a function of time for both the direct ionizing radiation and the UV reemission in the mutationally relevant 200-320 nm region. Adopting these models, standard supernova rates, and the expression for the time between events of given fluence from Scalo et al. (2003), we conclude that the surface of Mars is likely to have been exposed to bursts of ionizing radiation strong enough to sterilize or hypermutate most terrestrial eukaryotes about 1000 times from supernovae, over 100 times from γ-ray bursts, and over millions of times from stellar flares in the last Gyr (see Sec. 4 below for a discussion of the latter number). These numbers would be lower by roughly a factor of 10 for prokaryotes on average, although there are large variations in irradiation sensitivity in both domains.

## 4. Discussion: Astrobiological Importance of Redistributed UV

The deposited energy from planetary irradiation by very high energy astrophysical radiation sources is not completely lost in a planetary atmosphere, but rather a surprisingly large fraction can be redistributed by ionization and excitation into UV and visible radiation, even for thick atmospheres. This implies that extrasolar planets harboring life may be shielded from the incident X-rays and γ-rays from stellar flares, supernovae, and γ-ray bursts while still being subjected to highly stochastic UV and visible radiation environments over timescales ranging from minutes to eons.

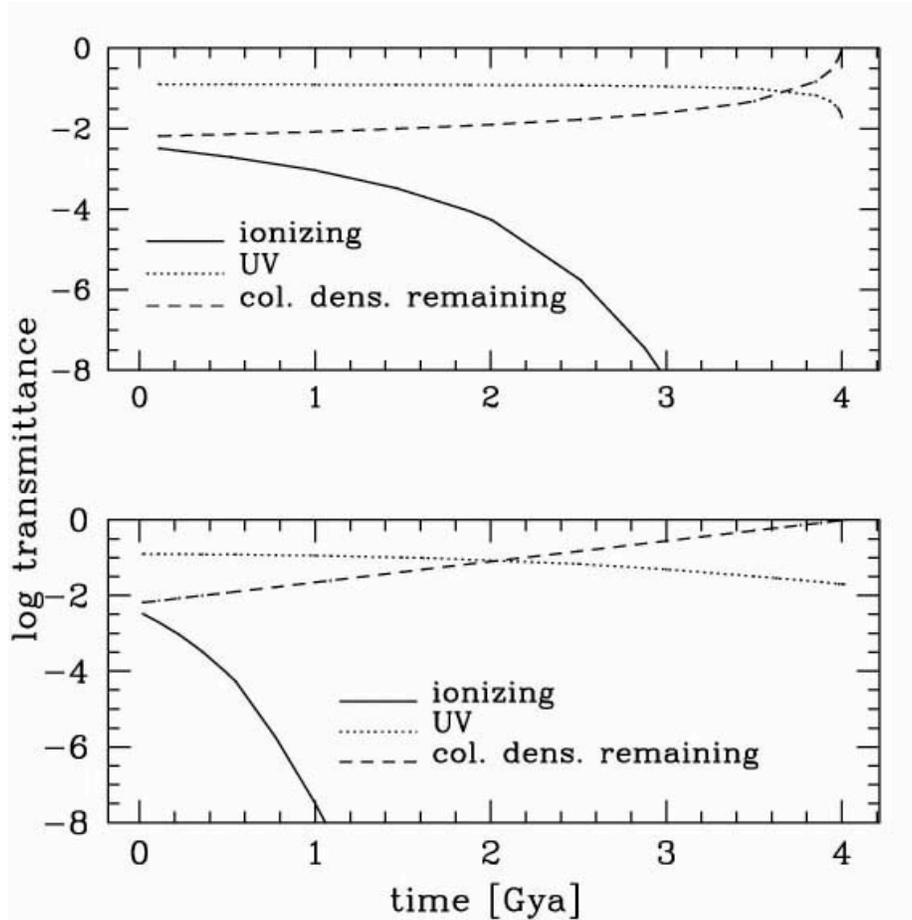

**Figure 2**:  Fraction of incident energy reaching the surface of Mars as ionizing and UV radiation as a function of time in the past for two simple models of the evolution of the $CO_2$ pressure in the Martian atmosphere, with present time on the left.  Upper and lower panels correspond to a surface pressure (and total column density) which decreases with time exponentially or as a power law, respectively.  The dashed lines represent the fraction of the initial atmospheric column density remaining as a function of time.  The initial column density was assumed to be 2600 g cm$^{-2}$, corresponding to a surface pressure of 1 bar.

The question of the conditions under which such intermittent ionizing radiation events are important is intricate, but can be approached by estimating the times between events for which the flux of the incident radiation redistributed to biologically active UV that we calculated for a column density equal to the Earth's exceeds the biologically significant UV background flux from the parent star of a conventional habitable zone planet, for which the total stellar flux is fixed at the solar constant within a factor of two or so, attenuated by whatever atmospheric UV screen is present. We postpone a detailed analysis of the range of extrasolar planets where the redistributed UV dominates that of the parent star and instead discuss a few key considerations here.

First consider the case of the Sun and Earth, and ignore ozone, so the discussion is applicable to the Archean Earth period or to similar star-planet pairs in which there is no strong UV shield. As mentioned above, there are large uncertainties in the Archean surficial flux (Cockell 2002), but recent work on sulfur isotope fractionation (Farquhar et al. 2002) suggests that the absence of a strong UV shield may be reasonable. The solar UV flux at 1 AU in the biologically active 200-320 nm region is about $1 \times 10^4$ erg cm$^{-2}$ s$^{-1}$, so, allowing for a 30% transmission due to Rayleigh scattering, a redistribution fraction of 4% implies that the incident ionizing flux due to, say, a flare must exceed about $1 \times 10^5$ erg cm$^{-2}$ s$^{-1}$ in order to exceed the Archean steady solar UV. Fluctuations in the UVB and UVC can be measured for the present Sun (Lean 1997, Lean 2001 for an update), and even reconstructed over the past 500 years (see Rozema et al. 2002 for a summary), and are known to be less than a few percent, so we can probably neglect variations in the direct solar UV, even in the Archean era.

The most energetic solar flares ever observed have energies of a few times $10^{32}$ erg. Assuming a duration of 10 min, this gives a flux of only 60 erg cm$^{-2}$ s$^{-1}$, far below the required threshold. The total observation time amounts to only a few decades, so it is possible that more energetic flares have occurred but have not yet been detected. The excellent agreement of the frequency-energy relations derived for energetic flares observed in hard X-rays by Crosby, Aschwanden, and Dennis (1993, SMM data for 2878 flares down to an energy of 25 keV) and Lin, Feffer, and Schwartz (2001, CGRO data for 167 flares down to 8 keV), as well as many other studies at lower energies (see Aschwanden et al. 2000) allows an extrapolation to higher energies. Using the data presented in these papers, it can be shown that the mean time between events of energy greater than $E_{32}$ in units of $10^{32}$ erg or of flux greater than F (cgs units, for 10 min duration) is given by

$$T = 5.6 \times 10^{-2} E_{32}^{0.6} = 4.8 \times 10^{-3} F^{0.6} \text{ yr.} \qquad (2)$$

where the index 0.6 of the power law is uncertain by about 0.2 according to data in the above-cited papers. So the solar UV would be exceeded by redistributed UV roughly once per decade *if* there exist solar flares with energy releases as large as $10^{36}$ erg. (This frequency should be decreased if the steeper soft X-ray energy-frequency distribution recently found by Veronig et al. (2002) is correct; in that case the exponent in eq. 2 would be 1.1.)

Schaeffer, King, and Deliyannis (2000) have identified nine cases of "superflares" with energy outputs from $10^{33}$ to $10^{38}$ erg on normal F8 to G8 main sequence stars. These flares cannot be attributed to binaries, rapid rotation, or youth, and therefore may be common in solar-type stars, although with recurrence times so long that they are not easily detected. Still, a large fluctuation in the surface UV for 10 minutes or so every decade or century is unlikely to have a significant biological effect unless it greatly pertubs atmospheric chemistry.

We contrast this with the situation for habitable zone planets orbiting lower-mass main sequence stars of spectral type M, which have masses between about 0.1 and 0.8 solar masses. Most stars in the Galaxy are low-mass red dwarf M (dM) stars (see Chabrier 2001 and references therein). It was previously suspected that tidal locking of habitable zone planets orbiting such stars would lead to atmospheric freeze-out on the "dark side," but climate simulations (Joshi, Haberle, and Reynolds 1997) demonstrate convincingly that atmospheric circulation is sufficient for retention of an atmosphere with pressures as low as 30 mbar, despite the expected tidal synchronization; even oceans are possible at larger pressures. Thus habitable planets orbiting very low-mass stars may be the most common abode for life.

A large fraction (perhaps half, Shakhovskaya 1994) of such stars are known as "flare stars" or "emission line stars" or "UV Ceti stars" (named after the prototype), usually designated as spectral type dMe, and are believed to spend perhaps the first several billion years of their lives in a state of intense flare activity, as do solar mass stars for shorter periods of time during their early evolution (see Gershberg et al. 1999 for access to an extensive database and bibliography).. These low mass stars are sources of intense flares with energies as large as $3 \times 10^{34}$ erg in ionizing radiation (e.g., Cully et al. 1993; Ramaty and Mandzhavidze 2000) roughly once per 100 hours of monitoring, with larger frequencies at smaller energies, similar to the frequency-energy relation for solar flares but shifted to higher energies and frequencies (see Gershberg and Shakhovskaya 1983 for a study of the flare energy-frequency scaling for about 20 dMe stars in the U and B photometric bands).

For stars of mass 0.6 and 0.2 $M_o$, the habitable zone radii are about 0.15 and 0.05 AU (A. Andreeshchev, J. Scalo, and D. Smith 2003, in preparation, hereafter ASS). As an example, the strong soft X-ray flares of energy $10^{34}$ erg with duration over 2 hours described by Cully et al. for the dMe flare star AU Mic would result in a flux above the planetary atmosphere of about $10^5$ erg cm$^{-2}$ s$^{-1}$, four percent of which should be redistributed at the planetary surface as UV according our results for the Archean Earth (Table 1), or 0.2 percent with a UV shield as opaque as Earth's ozone layer (Table 2). On the other hand the U-band (~210 to 380 nm Johnson photometric band) luminosities of dMe stars for these masses are less than the Sun by factors of about 0.05 and 0.02 (ASS). If the same fraction applies to shorter wavelengths, the steady parent star UV flux above the atmosphere would only be of order 200 erg cm$^{-2}$ s$^{-1}$. Clearly the redistributed flare flux at the surface, even in the presence of a strong UV shield, will greatly exceed this flux. Considering that smaller flare energies are more common, one expects the UV radiation environment of very low-mass stars to be completely dominated by redistributed flare energy occurring roughly once per day! In fact it is possible to show (ASS) that the planetary redistribution of even the relatively steady coronal X-rays could exceed the stellar UV radiation for these stars. In contrast, the solar coronal X-ray flux is negligible as a UV redistribution source, even when the Sun was younger and more active (see Ayres 1997, Güdel et al. 1997, Guinan et al. 2002 for reconstructions of early solar activity using solar proxies of different ages).

The above exercise shows that redistributed flare energy is probably a relatively marginal effect for habitable zone planets orbiting solar-like stars, but rapidly increases in importance for lower-mass stars, completely dominating the stellar UV for masses below 0.6-0.7 $M_o$ (K-type main sequence stars and cooler) if there is no ozone shield, and dominating even for an ozone shield like the Earth's for lower-mass dMe stars.

In the case of a UV shield, the redistributed biologically significant UV flux reaching the planetary surface will depend sensitively on the wavelength-dependence of both the UV shield and the absorption properties of the molecules specific to the planetary biochemistry. On the Earth, the ozone UV screen is highly effective, but strongly peaked at 260 nm. We expect an aerosol screen to absorb quite differently, with a relatively flat opacity (depending on the size of the particles). The action spectrum for many biological processes, especially various sorts of DNA damage, is also peaked at about 260 nm, with another, shorter-wavelength peak at about 200 nm, because of the wavelength dependence of the absorption cross sections of the specific purines and pyrimidines bases used in DNA (see Kolb, Dworkin, and Miller 1994, Fig. 7). It is important to bear in mind that, although extraterrestrial organisms may be coding in the sense of a genetic code and may even use base pairing for this code, it is possible or likely that the specific bases will be different from those used in DNA, with different spectral responses. There is a large literature concerning the possibility that the bases in DNA have changed, perhaps because the present bases would have been too unstable if temperatures were large in the habitat of the earliest life forms (e.g., Levy and Miller 1998). For example, the alternative bases urazole and guanazole proposed by Kolb, Dworkin, and Miller (1994) only absorb at the short wavelength peak of the present DNA bases. For more discussion see also Osawa (1995).

The additional consideration that makes the process discussed here unique in this regard is that the spectrum of redistributed UV should be relatively flat in the UV (like aurorae), rather than increasing strongly with increasing wavelength, as for stellar photospheric spectra. The solar irradiance spectrum measured at the Earth's surface is an example (A. Bais, presented in Rettberg and Rothschild 2000). The point is that it is the product of the source spectrum, the UV shield opacity, and the action spectrum for the biochemical effects of interest (e.g., mutations to coding organisms) that is important, and that this product is likely to vary strongly among inhabited habitable-zone planets.

We have not discussed more "exotic" potential sources of ionizing radiation such as supernova radioactive decay γ-rays (Scalo et al. 2003) and γ-ray bursts (Scalo and Wheeler 2002). Many of the same considerations apply to these sources, but the frequency with which any planet would be subjected to these sources is much smaller than the frequency of parent star flares. The case of supernova radioactivities is especially interesting because the duration of the event is much longer (several weeks) than a typical flare, opening up the possibility for real time evolutionary effects in small populations of bacteria-like organisms with small generation times, similar to the *in vitro* experiments of Ewing (1995, 1997). We also want to emphasize the long (but uncertain) durations of the subsequent cosmic-ray showers and attendant fluxes of γ-rays and other ionizing radiation, whose potential evolutionary significance was appreciated by Sagan and Shklovskii (1966).

## 5. Conclusions

Based on the detailed Monte Carlo calculations described above, we are able to make a number of conclusions concerning the radiation environment at the surface of terrestrial-like exoplanets.

1. Exoplanets with thick atmospheres ($\sim 1000$ g cm$^{-2}$) containing no abundant molecules that absorb in the biologically effective 200-320 nm region will receive at the ground of order 4% of the energy from an incident burst of ionizing radiation, independent of the form of the incident spectrum, as aurora-like ultraviolet reemission.

2. Exoplanets with thick atmospheres that contain an Earthlike altitude distribution and concentration of oxygen and ozone will receive of order 0.2% of the energy from incident bursts of ionizing radiation as ultraviolet reemission, although this number depends on the adopted upper limit for the biologically significant UV wavelength range.

3. Mars was shielded from incident $\gamma$-ray radiation only very early in its history, but even then it was subjected to the ultraviolet reemission processes discussed above that occur in planets with thick atmospheres. Today Mars is relatively unprotected from supernovae and $\gamma$-ray bursts, which characteristically have $\gamma$-ray spectra, but is well shielded from stellar flares, which have most of their energy in the lower-energy X-ray region of the spectrum.

4. Redistributed UV radiation due to ionizing radiation events will exceed the flux from the parent star at the surfaces of planets in the conventional liquid water habitable zone only occasionally for solar-type stars unless they are young, but should be the dominant source of UV flux for planetary systems orbiting the most common stars in our Galaxy, very low-mass stars.

These results suggest that surface and shallow-water coding organisms on such planets (and on the early Earth) may evolve very differently than on contemporary Earth, with diversity and evolutionary rate controlled by a stochastically varying mutation rate and frequent hypermutation episodes. We are currently investigating mathematical models that can guide our thinking about such relatively unexplored realms of evolutionary possibilities. In particular, it is completely unknown whether the predicted rate of evolution (and consequent development of cellular biochemical pathways and phenotypical complexity) under such circumstances should be accelerated or suppressed compared to evolution with a relatively steady mutation rate.

## Acknowledgments

We thank Peter Höflich for a suggestion that led to this work, Jim Kasting and Alex Pavlov for helpful communications, and two anonymous referees for constructive comments that improved the paper. DSS thanks the Harrington Doctoral Fellows Program and the NSF Graduate Research Fellowship Program for support. This research was supported by NSF grant 9907582.